\documentclass[a4paper,11pt]{article}
\pdfoutput=1 
\usepackage{jheppub} 
\usepackage[]{epsfig}
\usepackage{amsfonts}
\usepackage{amsmath}
\usepackage{subfigure}
\usepackage{graphicx}
\usepackage{graphics}
\usepackage{bbold}

\def\ba{\begin{eqnarray}}
\def\ea{\end{eqnarray}}
\def\be{\begin{equation}}
\def\ee{\end{equation}}

\title{Phase transitions in a holographic multi-Weyl semimetal}

\author[a]{Vladimir Juri\v ci\' c}
\author[b]{Ignacio Salazar Landea}
\author[c]{Rodrigo Soto-Garrido}
\affiliation[a]{Nordita, KTH Royal Institute of Technology and Stockholm University, Roslagstullsbacken 23, 10691 Stockholm, Sweden}
\affiliation[b]{Instituto de F\'\i sica de La Plata - CONICET, C.C. 67, 1900 La Plata, Argentina}
\affiliation[c]{Facultad de F\'isica, Pontificia Universidad Cat\'olica de Chile, Vicu\~{n}a Mackenna 4860, Santiago, Chile}

\emailAdd{juricic@gmail.com}
\emailAdd{peznacho@gmail.com}
\emailAdd{rodrigo.sotog@gmail.com}

\abstract{
Topological phases of matter have recently attracted a rather notable attention in the community dealing with the holographic methods applied to strongly interacting condensed matter systems. In particular, holographic models for gapless Weyl and multi-Weyl semimetals, characterized on a lattice by the monopole-antimonopole defects of the Berry curvature in momentum space, were recently formulated. In this paper, motivated by the quest for finding topological holographic phases, we show that holographic model for multi-Weyl semimetals features a rather rich landscape of phases. In particular, it includes a novel phase which we dub {$xy$ nematic} condensate, stable at strong coupling, as we explicitly show by the free energy and the quasi-normal mode analyses.  Furthermore, we provide its characterization through the anomalous transport coefficients.  We hope that our findings will motivate future works exploring the holographic realizations of the topological phases. }

\begin{document}
\maketitle

%%%%%%%%%%%%%%%%%%%%%%%%%%%%%%%%%%%%%%%%%
\section{Introduction}
\label{sec:introduction}
%%%%%%%%%%%%%%%%%%%%%%%%%%%%%%%%%%%%%%%%%%%%%%%%%%%%%%%%%%%

The paradigm of topological states of matter is by now well established, where in contrast to the usual Landau's symmetry classification, the characterization of a phase is provided in terms of the topological invariants~\cite{qi2011,hasan2010}. 
The prime examples of fermionic gapped topological phases are the integer quantum Hall states~\cite{klitzing1980}, with the time-reversal symmetry broken, for instance, by a magnetic field. As such, they are characterized in terms of the Chern number of the occupied electronic bands ~\cite{thouless1982}, directly related to the quantized Hall conductance in these systems. As a consequence of the nontrivial electronic topology, these topological states feature topologically protected gapless chiral edge modes, providing an example of the generic property of a topological state of matter: the (condensed-matter) bulk-boundary correspondence.  

The notion of topological phases is also operative when time-reversal symmetry is preserved, both in two~\cite{Kane-Mele-PRL2005} and three spatial dimensions~\cite{Fu-Kane-PRL2007,Moore-Balents-PRB2007,Fu-Kane-PRB2006}. These systems behave as ordinary insulators with a band gap in the bulk, but, in addition, host (time-reversal symmetry) protected gapless conducting states on their edge or surface \cite{hasan2010,qi2011}. Topological insulators were theoretically proposed~\cite{bernevig2006} and experimentally realized in two-dimensional HgTe/CdTe heterostructures~\cite{konig2007}. Subsequently, they have also been experimentally discovered in different three-dimensional compounds, such as Bi$_{1-x}$Sb$_x$ \cite{hsieh2008}, Bi$_2$Se$_3$, Bi$_2$ Te$_3$, and Sb$_2$Te \cite{xia2009,zhang2009,chen2009}.  Finally, free (noninteracting) gapped fermionic topological states can be systematically classified in the so-called tenfold periodic table through the fundamental anti-unitary time-reversal and particle-hole 
symmetries~\cite{schnyder-2008,schnyder-2009,kitaev-2009}, the latter accounting for the topological superconducting states featuring Majorana modes at the boundary.

Gapless systems can also be topological with the Weyl semimetals (WSMs) as the celebrated representatives~\cite{armitage2018,burkov2018}. Due to the broken time-reversal or spatial inversion symmetry, in a WSM the conduction and valence bands touch at pairs of points, the so-called Weyl points, in momentum space. Nontrivial electronic properties of these systems arise due to the  pseudo-relativistic linear dispersion close to the nodal points, which constitute topological point-like defects in momentum space, monopoles and antimonopoles, with the charge $n=\pm1$, representing the source and the sink of the abelian Berry curvature associated with the electronic wavefunctions. As a consequence, the topological protected Fermi arcs surface states emerge as the hallmark feature of the nontrivial electronic topology in these systems. 
WSMs have been experimentally observed in several compounds, including TaAs \cite{xu2015a,lv2015,huang2015}, TaP \cite{xu2015b} and NbAs \cite{xu2015c}; see also a review~\cite{hasan2017}. Furthermore, WSMs realize the chiral anomaly, originally proposed in the context of high energy physics~\cite{adler1969,bell1969}, giving rise to the exotic magneto-transport phenomena~\cite{zyuzin2012,son2013,huang2015b}.

Topological phases of matter, mostly due to their universal aspects, have also attracted a rather notable attention in high-energy physics, particularly in the community dealing with AdS/CFT correspondence~\cite{Maldacena-1999b,Witten-1998,aharony2000}, and its applications to strongly coupled condensed matter systems~\cite{zaanen2015,hartnoll2018}. As far as the gapless topological systems are concerned, a holographic realization of (strongly coupled) WSMs has been proposed in Ref. \cite{landsteiner2015}. Their topological invariants were computed in Ref.~\cite{Liu2018a}, while the generalization of WSMs featuring the line-like topological defect in momentum space, the nodal line semimetal, was studied by holographic methods in Ref.~\cite{Liu2018b}. The notion of the usual WSMs can also be generalized to  multi-Weyl semimetals (mWSMs) with the monopole charge $|n|>1$, although constrained to be at most $|n|=3$ due to the crystalline symmetries~\cite{XuPRL2011,FangPRL2012,BohmYang2014}. Such a higher than unity monopole charge arises  as a consequence of the low-energy dispersion which is linear in only one direction \cite{ParkS2017,Sukhachov2017,Huang2017,Dantas2018,Lepori2018,nag2018,Sengupta2019,SotoGarrido-2020}, while in the remaining two directions it scales as $k_{\perp}^{|n|}$.

In the present work, motivated by these developments, we show that a holographic model, recently proposed for the description of multi-Weyl semimetals~\cite{Dantas:2019rgp}, possesses a rich landscape of phases, which were not studied before. In particular, it includes a nematic condensate, a special case of which is the mWSM phase, with the $U(1)$ rotational symmetry restored. More importantly, we find an extra phase, which we dub \emph{$xy$ nematic condensate}, which, is stable at strong coupling, as we explicitly show by the free energy and the quasi-normal mode (QNM) analyses. Furthermore, we provide its characterization through the anomalous transport coefficients. We hope that our findings will motivate future works, where, for instance, the effects of the backreaction will be included to further elucidate the nature of the {$xy$ nematic} condensate. 

This paper is organized as follows. To set the stage, we review the basic features of the mWSMs and the holographic model, introduced in Ref.~\cite{Dantas:2019rgp}, in Sections~\ref{sec:mwsm} and~\ref{holomodel}, respectively. In Section \ref{phasetrans}, we study the instability of this model towards new strongly coupled phases, including a nematic phase and an {$xy$ nematic} condensate phase. Section \ref{sec:ls} is dedicated to the analysis of the linear stability of the low-lying QNMs in the corresponding phases. In Section \ref{anomtrans} we study the effect of the phase transition on the anomalous transport. Finally, in Section \ref{sec:end}, we summarize our results and outline possible future directions.

%%%%%%%%%%%%%%%%%%%%%%%%%%%%%%%%%%%%%%%%%%%%%%%%%%%%%%%
\section{Multi Weyl Semimetals}
\label{sec:mwsm}
%%%%%%%%%%%%%%%%%%%%%%%%%%%%%%%%%%%%%%%%%%%%%%%%%%%%%%%%%%

The minimal low-energy Hamiltonian of a mWSM with broken time-reversal symmetry is given by \cite{roy2017}
\begin{align}\label{eq:MWSM_Hamiltonian}
H^{(n)}_\pm({\bf k})&=\alpha_nk^n_\bot\left[\tau_x\cos(n\phi_k)+\tau_y\sin(n\phi_k) \right] \pm v_zk_z\tau_z
\end{align}
where  $k_\bot^2=k_x^2+k_y^2$, $\phi_k=\arctan\left(k_y/k_x \right)$, $\pm n$ is the monopole  charge  of  the two ($\pm$) Weyl nodes, and $\boldsymbol{\tau}$ are the Pauli matrices. This effective Hamiltonian can be obtained in the low-energy limit of a model with $n$ coupled simple Weyl fermion flavors~\cite{Dantas:2019rgp}
\begin{align}\label{eq:MWSM_HamiltonianDantas}
H^{(n)}_{\text{coup}}&=\left[v_\perp(k_x\tau_x+k_y\tau_y)+v_zk_z\tau_z\right]\otimes\mathbb{1}_{n\times n}+\Delta\left(\tau_x\otimes s_x^n+\tau_y\otimes s_y^n \right)
\end{align}
where $\mathbb{1}_{n\times n}$, $s_x^n$  and $s_y^n$ act  on  the  flavor  index, and   the $\mathbf{\tau}$'s act  on  the  pseudospin  index. Here, $\Delta$ term describes the  coupling among the single Weyl fermions, while $s^n$'s are the generators of the $n-$dimensional [the spin $(n-1)/2$] representation of the SU(2) group. 

The Lagrangian associated with the above Hamiltonian is given by
 \begin{equation}
 \label{freef}
 \mathcal{L}_{L} = i \psi^{\dagger}_{L} \tau^{\mu} \left[ \partial_{\mu} - i \Delta\left( \delta_{\mu}^x s_{x} +\delta_{\mu }^y s_{y}\right) \right] \psi_{L},
 \end{equation}
where $\tau^\mu=(\mathbb{1},\boldsymbol{\tau})$. Let us notice that the above Lagrangian can be conveniently rewritten in terms of a static non-Abelian gauge field as
\begin{equation}
 \label{eq:DWL_Interaction_2}
 \mathcal{L}_{L} = i \psi^{\dagger}_{L} \tau^{\mu} \left[ \partial_{\mu} - i A^a_{\mu} s_a\right] \psi_{L},
 \end{equation}\noindent
where $s_a = (s_0,s_i)$, with $i=x,y,z$, are the generators of $U(1)_L\times SU(2)_L$ and 
$A^a_{\mu}= \Delta (\delta_\mu^x \; \delta^{a}_x+ \delta_\mu^y \; \delta^{a}_ y)$.
The above Lagrangian represents a model for mWSMs with a non-Abelian $U(2)_L$ flavor symmetry, in the presence of a (non-Abelian) background gauge field $A_{\mu}^as_a$, with the form that reduces the  initial $SO(3,1)\times SU(2)_L$ symmetry group of $n$ decoupled copies of simple Weyl fermions to the diagonal $SO(1,1)\times U(1)_{3_L}$ symmetry corresponding to the mWSM. 
One can then  apply the standard field-theoretical techniques \cite{neiman2011,Landsteiner:2016led} to compute the non-Abelian anomalies, further supported by a corresponding holographic model~\cite{Dantas:2019rgp}.

Before we move on to consider possible phases in this  holographic model, notice that the Lagrangian (\ref{freef}) is invariant under a spatial rotation in the $x-y$ plane by an angle $\theta$ accompanied by an internal rotation $e^{-i \tau_3 \theta}$. In particular, the fermion bilinear  $ \psi^{\dagger}_{L} \tau^\mu\left(\delta_{\mu}^x s_{x} +\delta_{\mu }^y s_{y}\right)\psi_{L}$, realizing the mWSM phase in the dual field theory of the holographic model, is a scalar under this symmetry. { Other fermion bilinears related to the phases obtained in the holographic model break rotational symmetry. One of these corresponds to $\langle \psi^{\dagger}_{L} \tau^\mu \delta_{\mu}^x s_{x}\psi_{L}\rangle \neq \langle\psi^{\dagger}_{L}\tau^\mu \delta_{\mu }^y s_{y}\psi_{L} \rangle $, which we call  \emph{nematic} phase. 
%Finally, in another possible rotational symmetry breaking condensate  $\langle \psi^{\dagger}_{L} \tau^\mu \delta_{\mu}^y s_{x}\psi_{L}\rangle =\langle\psi^{\dagger}_{L}\tau^\mu \delta_{\mu }^x s_{y}\psi_{L} \rangle\neq 0 $ but also   $\langle \psi^{\dagger}_{L} \tau^\mu \delta_{\mu}^x s_{x}\psi_{L}\rangle = \langle\psi^{\dagger}_{L}\tau^\mu \delta_{\mu }^y s_{y}\psi_{L} \rangle\neq 0 $. 
As we will show explicitly in the next section, the holographic model features yet another phase
\begin{equation}
\label{axialbi2}
 \langle \psi^{\dagger}_{L} \tau^\mu\left(\delta_{\mu}^x s_{y} +\delta_{\mu }^y s_{x}\right)\psi_{L}\rangle\neq0, 
\end{equation}
which explicitly breaks the full rotational symmetry but preserves a combination of the $C_4$ rotation and the mirror $M_y$ operation, $C_4 M_y$. This combined operation is equivalent to the exchange of the $x$ and $y$ coordinates and we therefore call this phase 
\emph{$xy$ nematic} condensate. 

}

We would like to emphasize that, as shown in the next section, the {$xy$ nematic} condensate, as well as the nematic and the WSM phases, emerge as particular solutions within a rather general ansatz in the holographic model. 

%%%%%%%%%%%%%%%%%%%%%%%%%%%%%%%%%%%%%%%%%%%%%%
\section{The holographic model}
\label{holomodel}
%%%%%%%%%%%%%%%%%%%%%%%%%%%%%%%%%%%%%%%%%%%%%%%%%%%%

We start by considering the mWSM holographic model  \cite{Dantas:2019rgp}.
We will work in the probe limit with a finite temperature background geometry which we choose to be the Schwarzschild-AdS black hole 
\begin{equation}
\mathrm d s^2 = \frac{1}{r^2}\left(-f(r)\mathrm dt^2 + \frac{1}{f(r)}\mathrm dr^2+ dx^2 + dy^2 + dz^2 \right)\,,
\end{equation}
with the horizon at $r_h=1$, and the blackening factor $f(r)=1-r^4$. In these units the Hawking temperature is given by $T=\pi^{-1}$.

The action of the holographic model reads
\begin{eqnarray}
 S= - \int \mathrm{Tr}\left[\frac{1}{2n}\mathcal  F\wedge {^\star \mathcal F} +  \frac{1}{2c(n)}G\wedge{^\star G}+\lambda\left(A\wedge\mathrm (d A)^2 + \frac{3}{2} A^3\wedge \mathrm d A + \frac{3}{5} A^5\right)\right],\label{eq:HoloAct}
\end{eqnarray}
where the gauge fields are defined as
\begin{equation}
\mathcal A=\mathcal A^{0}s_0\quad,\quad \mathbb{A}=\mathbb{A}^{i}s_i\quad,\quad  A =\mathcal A+\mathbb{A}.
\end{equation}
%and $s_a=(s_0,s_i)$ are the identity and $SU(2)$ generators $s_i=\tau_i/2$ with $\tau_i$ as the Pauli matrices. 
For the sake of simplicity, we will exclusively work with $n=2$, which implies $c(n)=1/2$.
The corresponding field strength associated to the gauge fields are 
\begin{equation}
\mathcal F=d\mathcal A\quad,\quad G=d\mathbb{A} - i \mathbb{A}^2\quad,\quad F = \mathcal F+ G\,.
\end{equation}
The Maxwell-Yang-Mills-Chern-Simons equations on the curved background read
\begin{eqnarray}
\mathcal \nabla_\mu \mathcal F^{\mu\nu} - 6\lambda\epsilon^{\nu\rho\alpha\beta\gamma}\mathrm{Tr}\left( F_{\rho\alpha}F_{\beta\gamma} \right) &=& 0, \\
\mathcal D_\mu G^{a,\mu\nu} - \frac{3}{2}\lambda\epsilon^{\nu\rho\alpha\beta\gamma}\mathrm{Tr}\left( s^aF_{\rho\alpha}F_{\beta\gamma} \right) &=& 0\,.
\end{eqnarray}
By evaluating the action on-shell and recognizing the Bardeen counterterm, we find $\lambda$ to be of the form \footnote{{We refer the reader to Ref.~\cite{Dantas:2019rgp} where a careful derivation on both the field theoretical and holographic side was presented.}}
\begin{equation}
\lambda=\frac{1}{24\pi^2}. 
\end{equation}

To describe mWSM we break $SO(3,1)\times SU(2)_L\times U(1)_L$ symmetry down to $SO(1,1)\times U(1)_{3_L}\times U(1)_L$ with a background gauge field and turn on a magnetic potential in the $z$ direction. This translates into the boundary condition
\begin{equation}\label{eq:boundcond}
 A(r_b) =   \Delta\left(s_x\mathrm dx +s_y\mathrm dy\right)  + x B s_0\mathrm d y\,.
\end{equation}
The magnetic field breaks rotational $SO(3)\to SO(2)$. { The specific value of $B$ is not relevant for the features of our model, and for the numerical calculations we fix its value to be $B=1$. To be precise, for the phase structure of the model (Section \ref{phasetrans}) it decouples from the background equations of motion. 
It appears in the QNM analysis (Section \ref{sec:ls}) but since the mode structure turns out to be qualitatively independent  of $B$, as we explicitly checked, we only take the value $B=1$. Finally, as expected, the anomalous transport strongly depends on $B$, but in Section \ref{anomtrans} we define the coefficients so that this dependence does not explicitly appear.}
On the other hand,  the $\Delta$ term breaks the remaining $SO(2)$ rotation symmetry in the $x-y$ plane as well as the $U(1)$ gauge symmetry generated by $s_3$, but it preserves a combination of the two \cite{Gubser:2008zu}, yielding what  we will call rotational symmetry in the following.

We now consider a more general ansatz allowing us to turn on the desired sources at the boundary 
\begin{eqnarray}~\label{eq:ansatz}
 A(r) =  \left(Q_{x(1)}(r) s_x\,+Q_{x(2)}(r) s_y\,\right) dx +\left( Q_{y(2)}(r)s_y+Q_{y(1)}(r)s_x \right)dy + x B\, s_0\, d y\,.
\end{eqnarray}
Besides obeying the corresponding equations of motion, these fields are further subject to the constraint
\begin{equation}
\label{constr}
    Q_{x(1)}Q_{x(2)}'-Q_{x(2)}Q_{x(1)}'+Q_{y(1)}Q_{y(2)}'-Q_{y(2)}Q_{y(1)}'=0.
\end{equation}
There are three simple solutions to this constraint:
\begin{itemize}
\item The mWSM (normal) phase: This is probably the simplest ansatz and corresponds to $Q_{x(2)}=Q_{y(1)}=0$ and $Q_{x(1)}=Q_{y(2)}=Q$~\cite{Dantas:2019rgp}.
    \item The nematic phase:  $Q_{x(2)}=Q_{y(1)}=0$ and $Q_{x(1)}\neq Q_{y(2)}$. This phase breaks the rotational invariance.
    \item The {$xy$ nematic} condensate:  $Q_{x(1)}=Q_{y(2)}=Q_1$ and $Q_{x(2)}= Q_{y(1)}=Q_2$. This phase { does not} preserve the rotational invariance and we associate it with the condensation of an operator as given by~(\ref{axialbi2}) in the dual field theory.
\end{itemize}

%%%%%%%%%%%%%%%%%%%%%%%%%%%%%%%%%%%%%%%%%%%%%%%%%%%%%%%
\section{The phase transitions}
\label{phasetrans}
%%%%%%%%%%%%%%%%%%%%%%%%%%%%%%%%%%%%%%%%%%%%%%%%%%%%%%%%%%%

Now that we have recognized, besides the mWSM, other two candidates for instabilities, we study the existence of the corresponding solutions with the appropriate boundary conditions (\ref{eq:boundcond}). We find that indeed both instabilities can occur for large enough $\Delta$ but the {$xy$ nematic} condensate is energetically more favorable. 

We notice that by virtue of this simple ansatz $\lambda$ disappears from the background equations of motion, implying that such an instability is universal to any Yang-Mills theory living in an AdS universe when turning on this $\Delta$ source. This instability is a close relative of the holographic p-wave superconductor  \cite{Gubser:2008wv,Zayas:2011dw}. More generally, it may be considered as a new member of the family of the Einstein-Yang-Mills instabilities in AdS spacetime~\cite{Winstanley:2008ac}.

%%%%%%%%%%%%%%%%%%%%%%%%%%%%%%%%%%%%%%%%
\subsection{The nematic phase}
%%%%%%%%%%%%%%%%%%%%%%%%%%%%%%%%%%%%%%%%%

Plugging the ansatz (\ref{eq:ansatz}) into the equations of motion we find 
\begin{align}
    \left(\frac{Q_x'f}{r}\right)'-\frac{Q_y^2}{r}Q_x=0 \,,\nonumber\\
    \left(\frac{Q_y'f}{r}\right)'-\frac{Q_x^2}{r}Q_y=0.
    \label{eombg}
\end{align}

\noindent We now proceed to integrate numerically Eq.~(\ref{eombg}) starting from the near horizon towards the boundary
\begin{align}
    Q_x\approx Q_{x\,h}+\frac14 Q_{y\,h}^2 Q_{x\,h} (1-r)+\dots\nonumber\\
    Q_y\approx Q_{y\,h}+\frac14 Q_{x\,h}^2 Q_{y\,h} (1-r)+\dots
\end{align}
 and we use the free horizon parameters to match the desired boundary conditions (\ref{eq:boundcond}). These boundary conditions can be satisfied by setting $Q_{x\,h}=Q_{y\,h}$. Then the equations of motion imply $Q_x=Q_y$ for all $r$. 
 
 Slightly more challenging is to find solutions where $Q_{x\,h}\neq Q_{y\,h}$ but still satisfy the boundary condition given by Eq.~(\ref{eq:boundcond}). We find nonetheless that such solutions do exist for large enough $\Delta$ and that they are energetically favorable with respect to the mWSM phase. This family of solutions is characterized by the UV expansion for the $Q_i$ in the form
\begin{align}
    Q_x\approx\Delta+{J_x^{(1)}}r^2+\dots\nonumber\\
    Q_y\approx\Delta+{J_y^{(2)}}r^2+\dots 
\end{align}
The spontaneous breaking of rotational symmetry can then be characterized by the difference between the expectation values associated with $Q_i$, $\langle O \rangle= J_x^{(1)}-J_y^{(2)}$, which we plot in Fig.~\ref{pt} after re-scaling by $\Delta^3$ (based on dimensional grounds). There we can identify the critical value of $\Delta_c\approx 7.92$ for the nematic instability to set in.

%%%%%%%%%%%%%%%%%%%%%%%%%%%%%%%%%%%%%%%%%%%%
\subsection{The $xy$ nematic condensate}
%%%%%%%%%%%%%%%%%%%%%%%%%%%%%%%%%%%%%%%%%%

We now turn to the analysis of the {$xy$ nematic} condensate phase. The equations of motion read
\begin{align}
    \left(\frac{Q_1'f}{r}\right)'+\left(Q_2^2-Q_1^2\right)\frac{Q_1}{r}=0\,,\nonumber \\
     \left(\frac{Q_2'f}{r}\right)'+\left(Q_1^2-Q_2^2\right)\frac{Q_2}{r}=0\,,
\end{align}
which we integrate from the near horizon
\begin{align}
    Q_1\approx Q_{1\,h}+\frac14 \left(Q_{1\,h}^2-Q_{2\,h}\right)Q_{1\,h}(1-r)+\dots \nonumber \\
    Q_2\approx Q_{2\,h}+\frac14 \left(Q_{2\,h}^2-Q_{1\,h}\right)Q_{2\,h}(1-r)+\dots
\end{align}
to match the desired boundary conditions 
\begin{align}
    Q_1\approx& \Delta+{J_1}r^2+\dots\nonumber\\
    Q_2\approx& {J_2}r^2+\dots
\end{align}

\begin{figure}[t!]
\begin{center}
\includegraphics[width=0.495\textwidth]{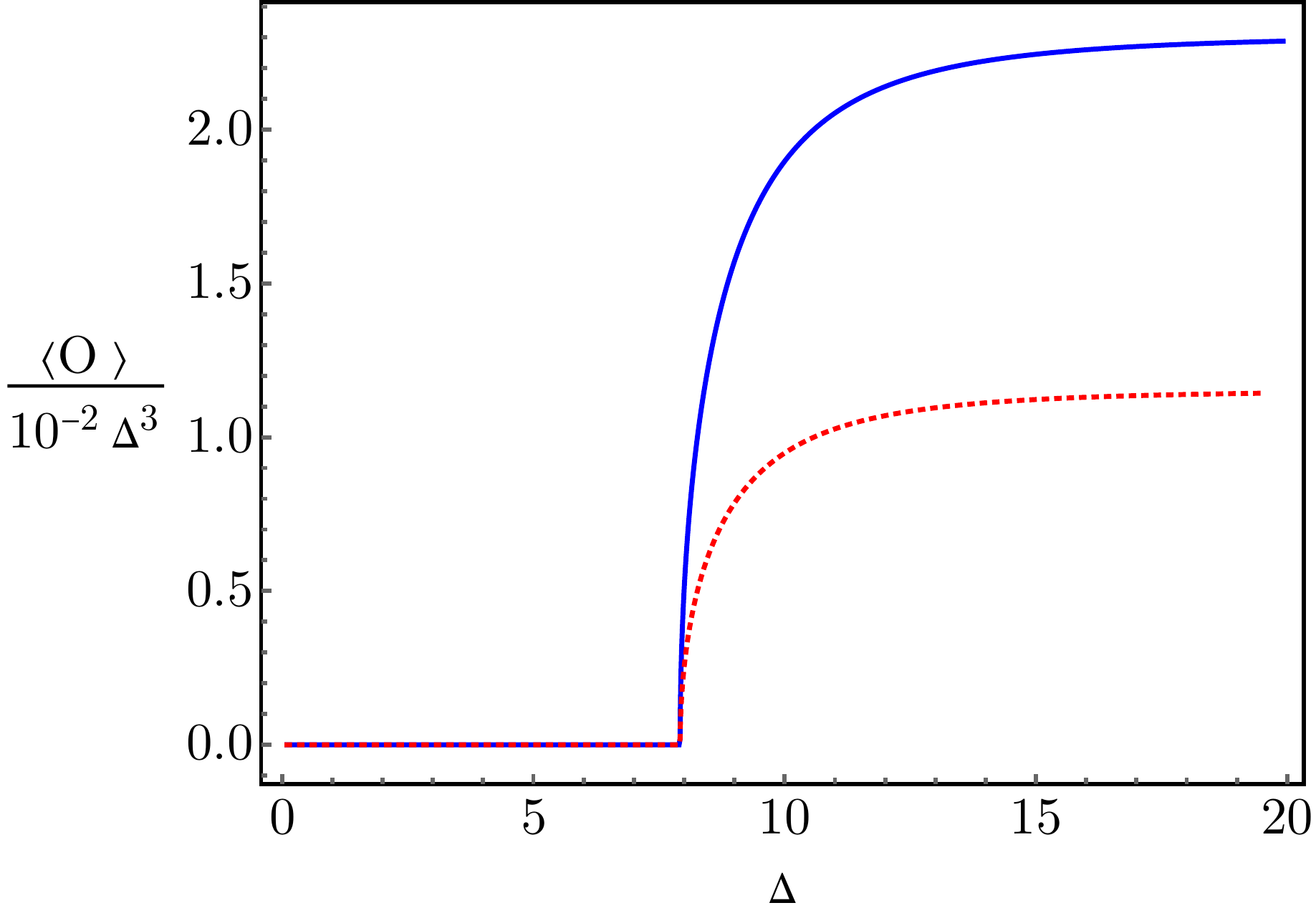}\hfill \includegraphics[width=0.505\textwidth]{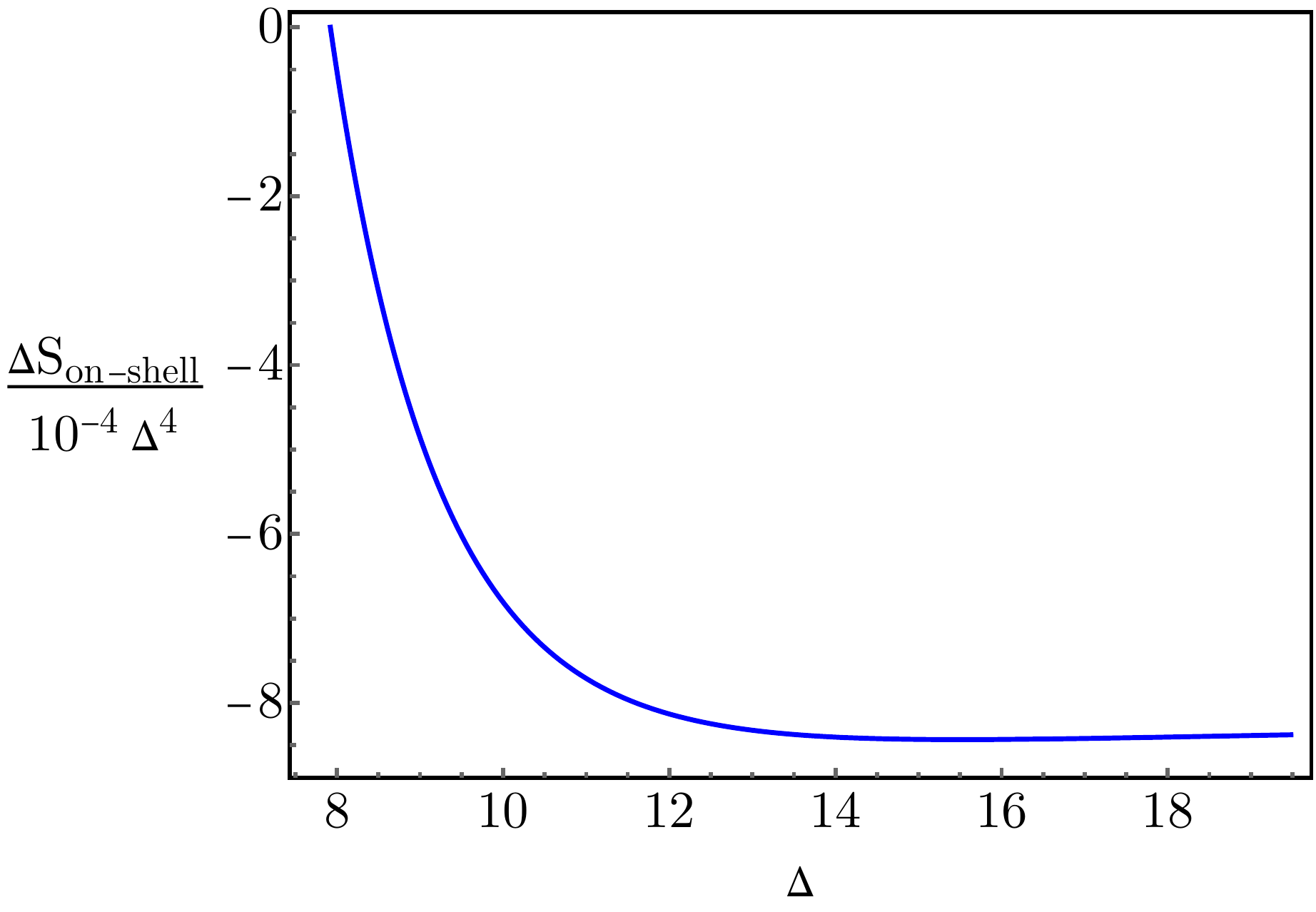}
\caption{\label{pt}Left: Order parameter  $\langle O\rangle/\Delta^3$ as a function of $\Delta$ for the nematic (solid blue) and {$xy$ nematic} condensate (dotted red) phases. Right: Difference of free energies between the {$xy$ nematic} condensate phase and the normal phase.}
\end{center}
\end{figure}

\noindent Again we find that the {$xy$ nematic} condensate exists for large enough $\Delta$ with the exact same critical value $\Delta_c\approx 7.92$ as  for the nematic phase. The order parameter corresponding to the new phase is $\langle O\rangle=J_2$. We plot this order parameter together with the one for the nematic phase in Fig.~\ref{pt}.
%%%%%%%%%%%%%%%%%%%%%%%%%%%%%%%%%%%%%%%%%%%%%%%%%%%%%
%%%%%%%%%%%%%%%%%%%%%%%%%%%%%%%%%%%%%%%%%%%%%%%%%%%
%%%%%%%%%%%%%%%%%%%%%%%%%%%%%%%%%%%%%%%%%%%%%%%%%%%
%%%%%%%%%%%%%%%%%%%%%%%%%%%%%%%%%%%%%%%%%%%%%%%%%%%%%%%%%%

\begin{figure}[t]
\begin{center}
\includegraphics[width=0.515\textwidth]{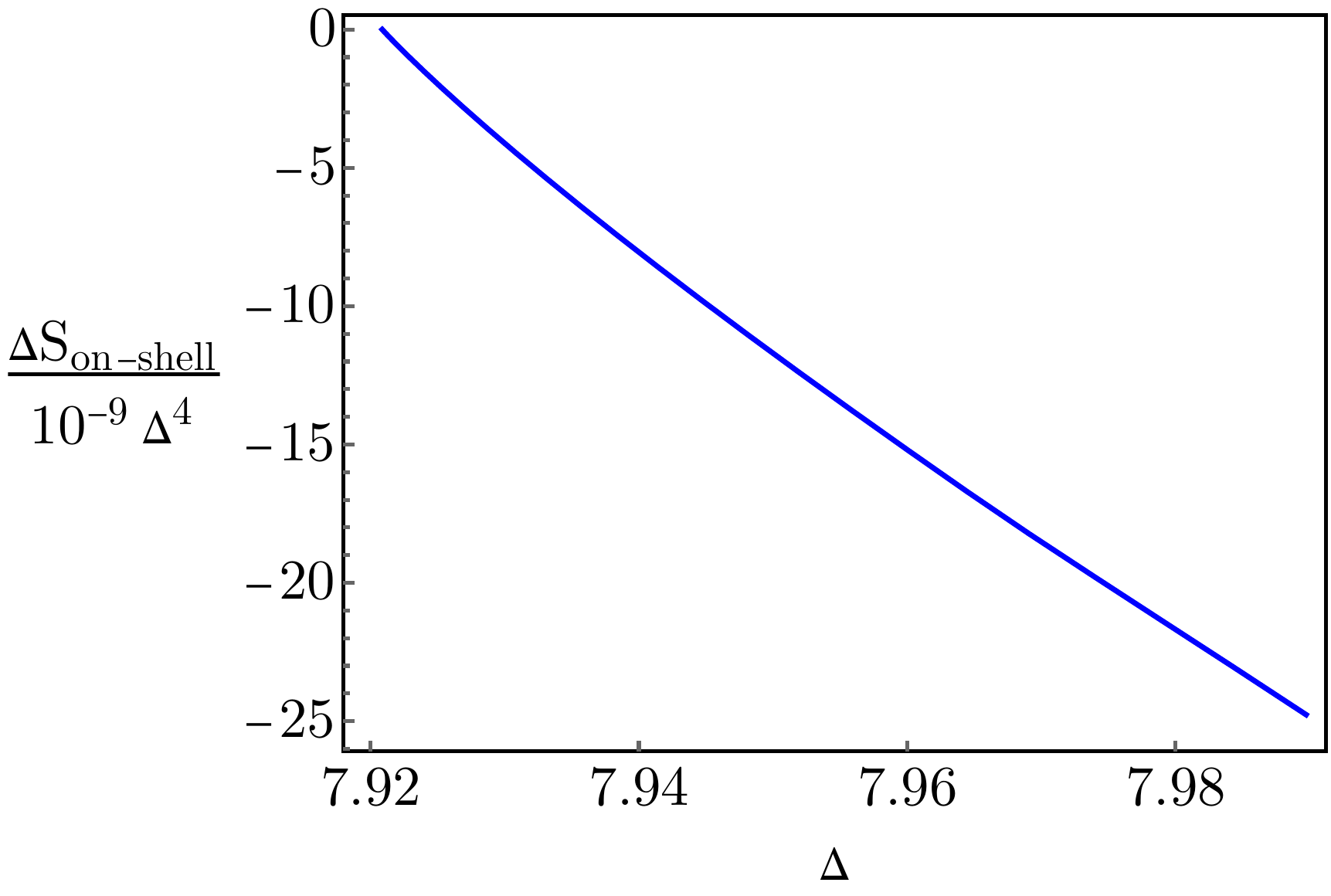}\hfill \includegraphics[width=0.485\textwidth]{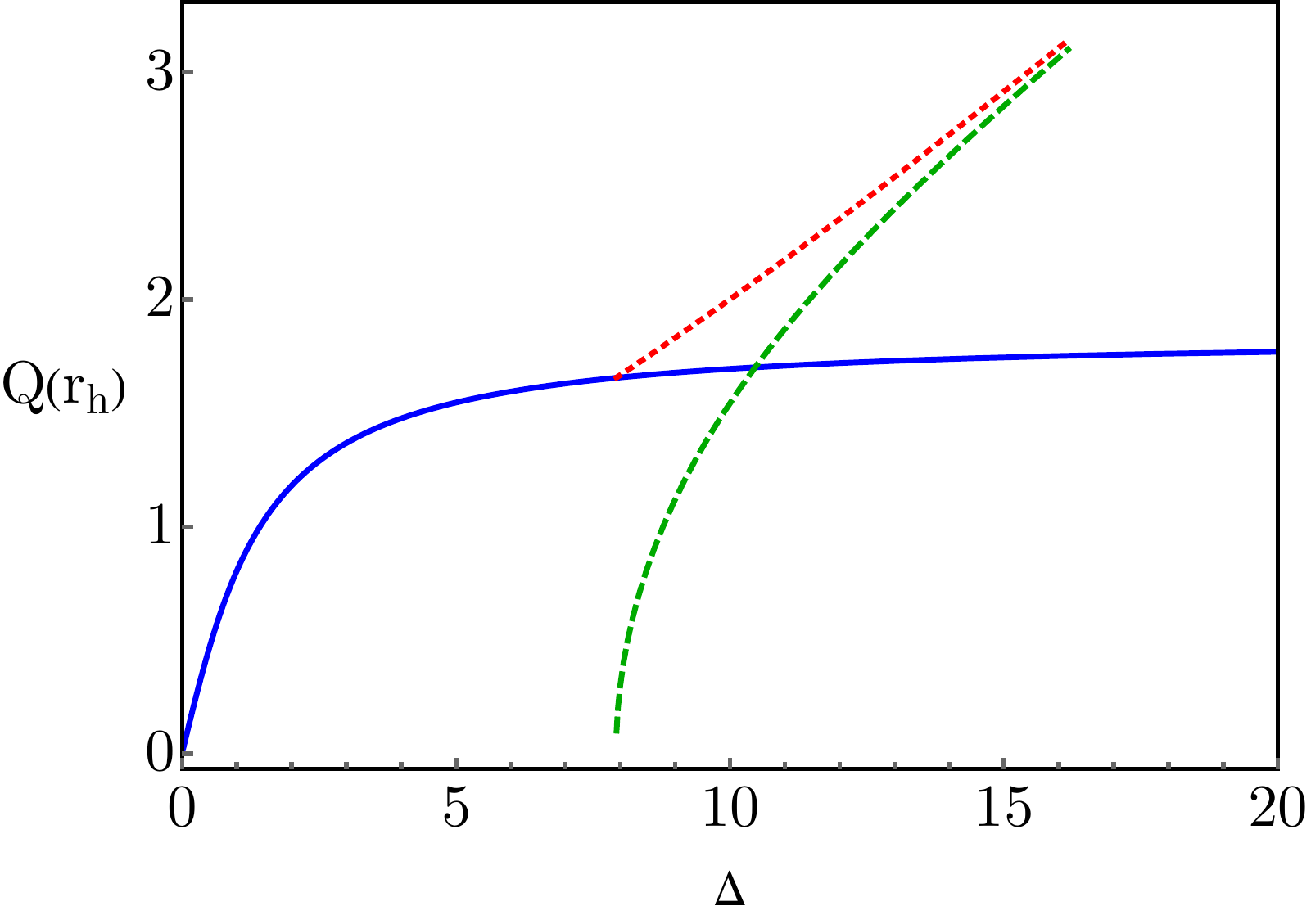}
\caption{\label{deltair}Left: Difference of free energies between the {$xy$ nematic} condensate phase and the nematic phase. Right: Horizon value for the different $Q_i$. The solid blue curve corresponds to the mWSM or normal phase $Q(r_h)$. We also plot $Q_1(r_h)$ (dotted red) and $Q_2(r_h)$ (dashed green) for the {$xy$ nematic} condensate phase. }
\end{center}
\end{figure}

As a first step to study the stability of these new solutions, we look for the on-shell action that corresponds to the free energy in the dual field theory. Explicitly evaluating the on shell action we find
\begin{equation}
    S_{on-shell}=-\Delta\left(J_{x(1)}+J_{y(2)}\right)+\int_\epsilon^{1-\epsilon}\frac{dr}r \left(Q_{x(2)}Q_{y(1)}-Q_{x(1)}Q_{y(2)}\right)^2.
\end{equation}
\noindent We regularize the above on-shell action to facilitate the stability analysis carried out by directly evaluating the difference between the on-shell actions for the normal and condensed phases. We find that the {$xy$ nematic} phase is stable while the nematic condensate is unstable for strong coupling $\Delta$, as shown in the right panel of Fig.~\ref{pt}.  The difference between them is very small, though, ${\mathcal O}(10^{-8})$, as shown in the left panel of Fig.~\ref{deltair}, making the computation numerically challenging.  Since the phase transition is of the second order (continuous), we can analyze the stability by invoking the linear perturbations. The corresponding results are discussed in Section \ref{sec:ls}, and agree with the above free energy computation.

In addition, we now study the RG flow of the relevant parameter $\Delta$. The horizon values for the parameters $Q_i$ corresponding to the mWSM and {{$xy$ nematic}} phases are displayed Fig.~\ref{deltair}. In the mWSM phase, we see that $Q(r_h)$  grows linearly for small $\Delta$ and then saturates to a constant at large values of this parameter. However, this phase becomes  unstable for some critical value of the parameter $\Delta$,  consistent with the results from the above free energy analysis. We also plot the would be values of $Q(r_h)$ at large $\Delta$ as if the instabilities did not take place in the normal phase. The horizon parameter $Q_1(r_h)$ for the {$xy$ nematic} phase at the critical $\Delta=\Delta_c$ departs from the behavior in the normal phase and grows linearly with $\Delta$, red curve in the right panel of Fig.~\ref{deltair}.
As we can also see from the right panel of Fig.~\ref{deltair}, $Q_2(r_h)$ (green curve) behaves as $(\Delta-\Delta_c)^{1/2}$ at small $\Delta$, while it approaches $Q_1(r_h)$ at large values of the parameter $\Delta$.

%%%%%%%%%%%%%%%%%%%%%%%%%%%%%%%%%%%%%%%%%%%%
%%%%%%%%%%%%%%%%%%%%%%%%%%%%%%%%%%%%%%%%%%%%
%%%%%%%%%%%%%%%%%%%%%%%%%%%%%%%%%%%%%%%%%%%%%%%%
%%%%%%%%%%%%%%%%%%%%%%%%%%%%%%%%%%%%%%%%%%%%%%%

%%%%%%%%%%%%%%%%%%%%%%%%%%%%%%%%%%%%%%%%%%%%%%%%%
%%%%%%%%%%%%%%%%%%%%%%%%%%%%%%%%%%%%%%%%%%%%%%%%
\section{Linear stability}
\label{sec:ls}
%%%%%%%%%%%%%%%%%%%%%%%%%%%%%%%%%%%%%%%%%%%%%%%
%%%%%%%%%%%%%%%%%%%%%%%%%%%%%%%%%%%%%%%%%%%%%%%

To further understand these new solutions, we study their stability under small perturbations by considering the spectrum of the low lying  QNMs. We perform a zero-momentum analysis following the approach used for the p-wave superfluid~\cite{Gubser:2008wv}. This analysis is important as a consistency check with the previous stability analysis based on the free energy. As we will see, the Chern-Simons terms make a large set of the fields coupled to each other. To deal with this, we employ the determinant method \cite{Amado:2009ts}  which can  give the poles of the Green functions without the need of decoupling the equations.

{ We proceed by solving the linearized time-dependent  equations of motion around the different background solutions previously found.} We focus on the mWSM and the {$xy$ nematic} condensate phases, since they are energetically favorable, as previously found, and all the plots are made for the stable backgrounds. Nevertheless, we did check that some poles of the Green function cross to the upper complex plane when one considers an energetically unfavorable phase, signaling, as expected, a tachyonic mode. To be precise, we will consider fluctuations on top of the {$xy$ nematic} condensate solution, emerging from the normal phase for $\Delta>\Delta_c$. 

We decouple this way obtained equations in different sectors by turning on a subset  of the fields. The retarded Green function are obtained by virtue of the infalling boundary conditions at the horizon
\begin{equation}
    a\approx (1-r)^{-i \omega/4}\left(a_h+O(1-r)\right)\,,
\end{equation}
where the other coefficients of the series are fixed in terms of the $a_h$'s.
Generically there are as many independent solutions as fields we need to turn on, fixed by the values for the different horizon coefficients $a_h$. The situation is different when we turn on a temporal component of the gauge field. Then regularity at the horizon forces the  corresponding $a_h$'s to be zero, and we can use a pure gauge solution to have a complete set of independent solutions.

In the following subsections we identify the relevant sectors and find a number of the lowest lying poles in the corresponding Green functions.

%%%%%%%%%%%%%%%%%%%%%%%%%%%%%%%%%%%%%%%%%%%%%%%%%%%
%%%%%%%%%%%%%%%%%%%%%%%%%%%%%%%%%%%%%%%%%%%%%%%%%%%
\subsection{The Higgs sector}
%%%%%%%%%%%%%%%%%%%%%%%%%%%%%%%%%%%%%%%%%%%%%%%%%
%%%%%%%%%%%%%%%%%%%%%%%%%%%%%%%%%%%%%%%%%%%%%%%%%%%%%

\begin{figure}[t!]
\begin{center}
\includegraphics[width=0.485\textwidth]{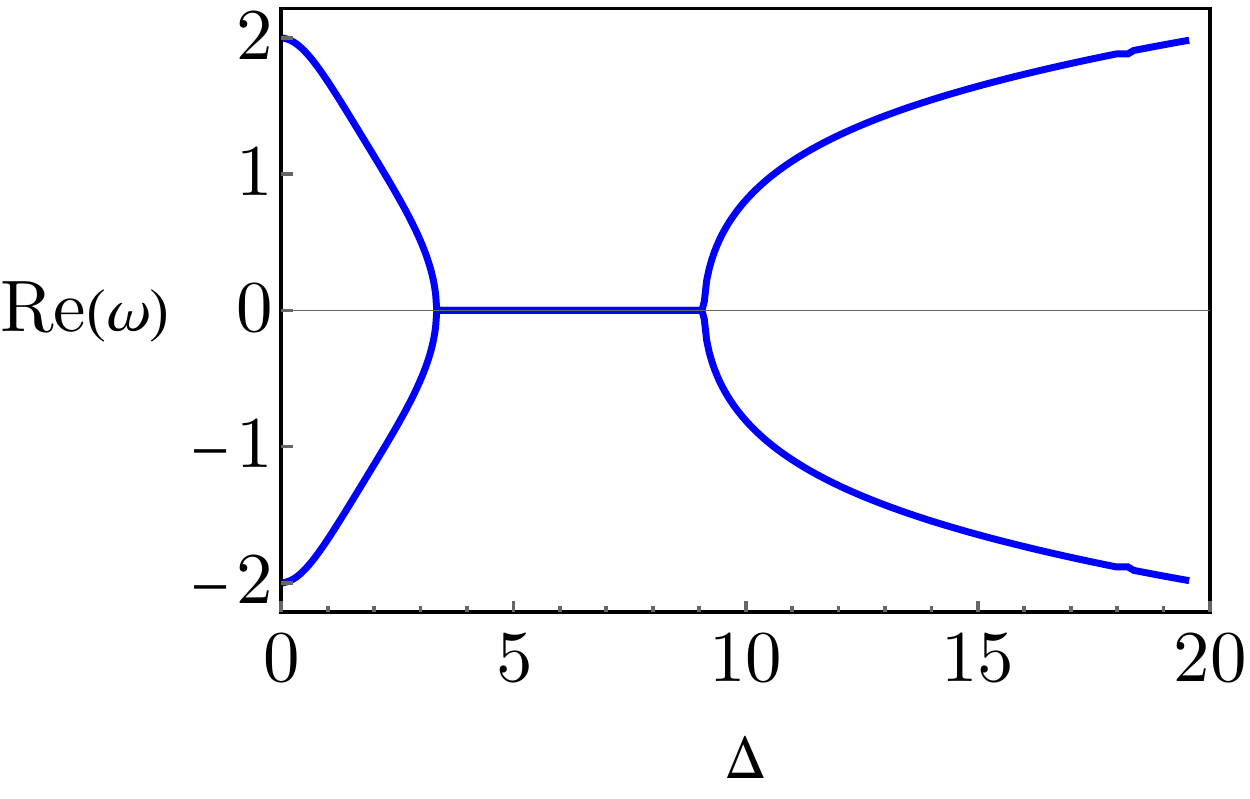}\hfill \includegraphics[width=0.51\textwidth]{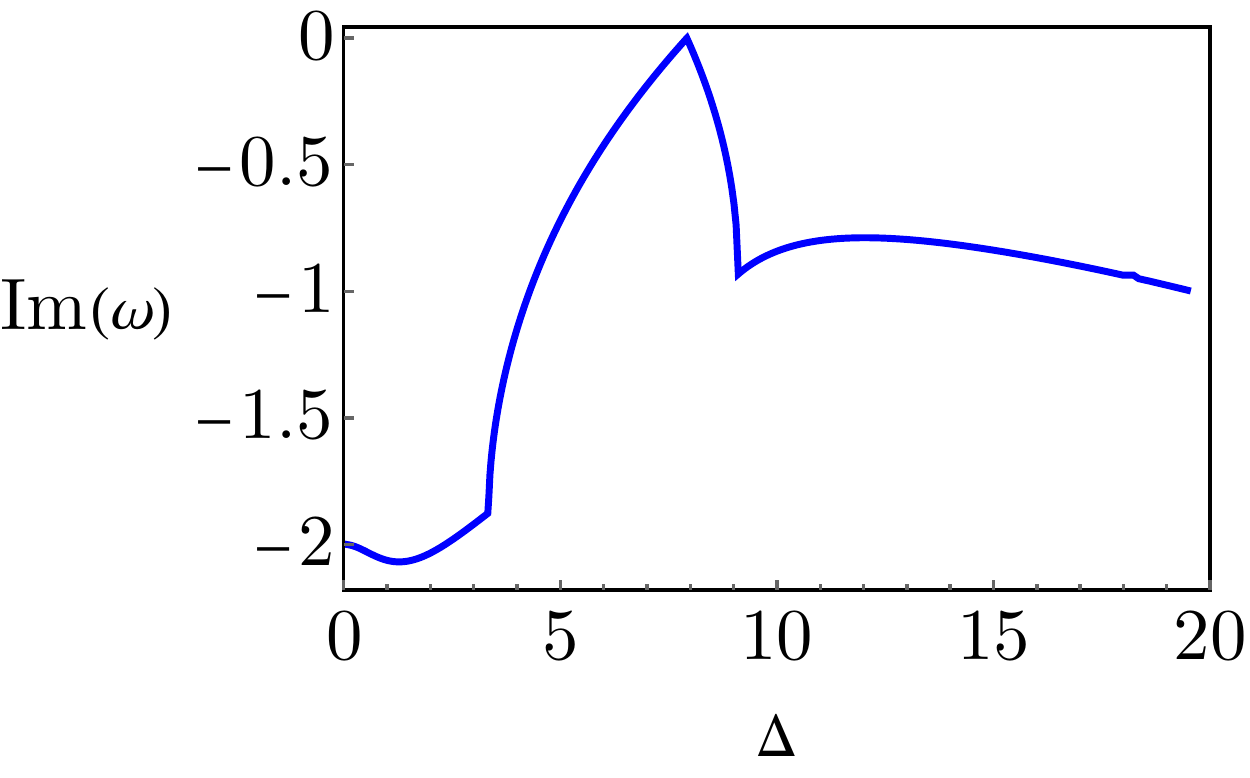}
\caption{\label{higgs} Real and imaginary part of the frequency of the lowest lying QNM as a function of the parameter $\Delta$.}
\end{center}
\end{figure}

Let us start by considering the fluctuations of the form
\begin{equation}
    \delta A= e^{-i\omega t} q_{+\bot}(r) \left(s_2dx+s_1dy\right) +e^{-i\omega t} q_{+\parallel}(r)\left(s_2dy+s_1dx\right),
\end{equation}
with the corresponding equations of motion given by
\begin{align}
    \left(\frac{q_{+\bot}'f}r\right)'+\left(\frac{\omega^2}{rf}-\frac{Q_1^2-3Q_2^2}{r}\right)q_{+\bot}+\frac{2Q_1 Q_2 q_{+\parallel}}{r}&=0\,,\nonumber\\
     \left(\frac{q_{+\parallel}'f}r\right)'+\left(\frac{\omega^2}{rf}-\frac{Q_2^2-3Q_1^2}{r}\right)q_{+\parallel}+\frac{2Q_1 Q_2 q_{+\bot}}{r}&=0.
\end{align}
Focusing on the lowest lying modes, we can track them as a function of $\Delta$, and the result is  displayed in Fig.~\ref{higgs}. As we can see, the imaginary part of the quasinormal frequencies has three spikes as a function of $\Delta$. The middle one corresponds to the pole hitting the real axis and bouncing back into the lower half plane. This happens exactly at the critical $\Delta_c$. If one had instead considered the mWSM solutions, these modes would have continued their journey to the upper half plane, signaling the instability towards this new phase.
%%%%%%%%%%%%%%%%%%%%%%%%%%%%%%%%%%%%%%%%%%%%%%%%%%%%
%%%%%%%%%%%%%%%%%%%%%%%%%%%%%%%%%%%%%%%%%%%%%%%%%%%%
%%%%%%%%%%%%%%%%%%%%%%%%%%%%%%%%%%%%%%%%%%%%%%%%%%%%%%%
%%%%%%%%%%%%%%%%%%%%%%%%%%%%%%%%%%%%%%%%%%%%%%%%%%%%
The other spikes in Im$(\omega)$ correspond to particular values of $\Delta$ where two complex conjugate frequencies collapse rendering the pole to be purely imaginary, which happens at $\Delta \approx 3.35$ and $\Delta \approx 9.10$.

The lowest lying QNM controls the behavior of the systems after a sudden quench \cite{Bhaseen:2012gg}. Based on this observation, one can infer that there might be a characteristic quantum critical region near the phase transition where the dynamical response of some operators change from being oscillatory to purely diffusive. This sector is also stable for the nematic solutions. However, this will not be the case for the Goldstone sector, which we turn to next.

%%%%%%%%%%%%%%%%%%%%%%%%%%%%%%%%%%%%%%%%%%%%%%%%%%%%%%%%
%%%%%%%%%%%%%%%%%%%%%%%%%%%%%%%%%%%%%%%%%%%%%%%%%%%%%%%
\subsection{The Goldstone sector}
%%%%%%%%%%%%%%%%%%%%%%%%%%%%%%%%%%%%%%%%%%%%%%%%%%%
%%%%%%%%%%%%%%%%%%%%%%%%%%%%%%%%%%%%%%%%%%%%%%%%%%%%%%%%

This sector arises  from turning on the perturbations of the form
\begin{align}
    \delta A=& e^{-i\omega t}(s_0 a_{t(0)}(r)+s_3 a_{t(3)}(r))dt+e^{-i\omega t}(s_0 a_{z(0)}(r)+s_3 a_{z(3)}(r))dz+\nonumber\\
    &+e^{-i\omega t} q_{-\bot}(r)\left(s_2 dx-s_1 dy\right)+e^{-i\omega t} q_{-\parallel}(r)\left(s_1 dx-s_2 dt\right).
\end{align}
The equations of motion read
\footnotesize
\begin{align}
   & \left( \frac{a_{t(0)}'}{r}\right)'+12 \lambda \left(Q_1^2 a_{z(3)}\right)'-12 \lambda \left(Q_2^2 a_{z(3)}\right)'=0\,, \nonumber\\
    & \left( \frac{a_{z(0)}'f}{r}\right)'+12 \lambda\left( \left(Q_1^2 a_{t(3)}\right)'-\left(Q_2^2 a_{t(3)}\right)'\right)+\frac{\omega^2a_{z(0)}}{rf}+24i\lambda\omega\left(q_{-\bot}Q_1'-q_{-\parallel}Q_2'\right)+96\lambda B a_{t(0)}'=0\,,\nonumber\\
     & \left( \frac{q_{-\parallel}'f}{r}\right)'-\frac{i \omega a_{t(3)}Q_2}{rf}+12i\lambda \omega a_{z(0)}Q_2'+\left(\frac{\omega^2}{rf}+\frac{Q_1^2-Q_2^2}{r}\right)q_{-\parallel}=0\,,\nonumber\\
     & \left( \frac{q_{-\bot}'f}{r}\right)'+\frac{i \omega a_{t(3)}Q_1}{rf}+12i\lambda \omega a_{z(0)}Q_1'+\left(\frac{\omega^2}{rf}-\frac{Q_1^2-Q_2^2}{r}\right)q_{-\bot}=0\,,\nonumber\\
     & \left( \frac{a_{t(3)}'}{r}\right)'+12\lambda \left( Q_1^2-Q_2^2\right)a_{z(0)}'-\frac{2 a_{t(3)}\left(Q_1^2+Q_2^2\right)}{rf}+\frac{2i\omega\left(Q_1 q_{-\bot}-Q_2 q_{-\parallel}\right)}{rf}+24\lambda B a_{z(3)}'=0 \,,\nonumber\\
     & \left( \frac{a_{z(3)}'f}{r}\right)'+12\lambda \left(Q_1^2-Q_2^2\right)a_{t(0)}'-24\lambda B a_{t(3)}'+\left(\frac{\omega^2}{rf}-2\frac{Q_1^2+Q_2^2}{r}\right)a_{z(3)}=0, 
\end{align}
\normalsize
which are subject to the constraints 
\begin{align}
    &24 i \lambda B \omega a_{z(3)}+12 i \lambda \omega \left(Q_1^2-Q_2^2\right)+\frac{i\omega a_{t(3)}'}{r}+\frac{2f}r\left(Q_1^2\left(\frac{q_{-\bot}}{Q_1}\right)'-Q_2^2\left(\frac{q_{-\parallel}}{Q_2}\right)'\right)=0\,,\nonumber\\
    &a_{t(0}'+12\lambda r \left(8 B a_{z(0)}+\left(Q_1^2-Q_2^2\right)a_{z(3)}\right)=0.
\end{align}

\begin{figure}[t!]
\begin{center}
\includegraphics[width=0.485\textwidth]{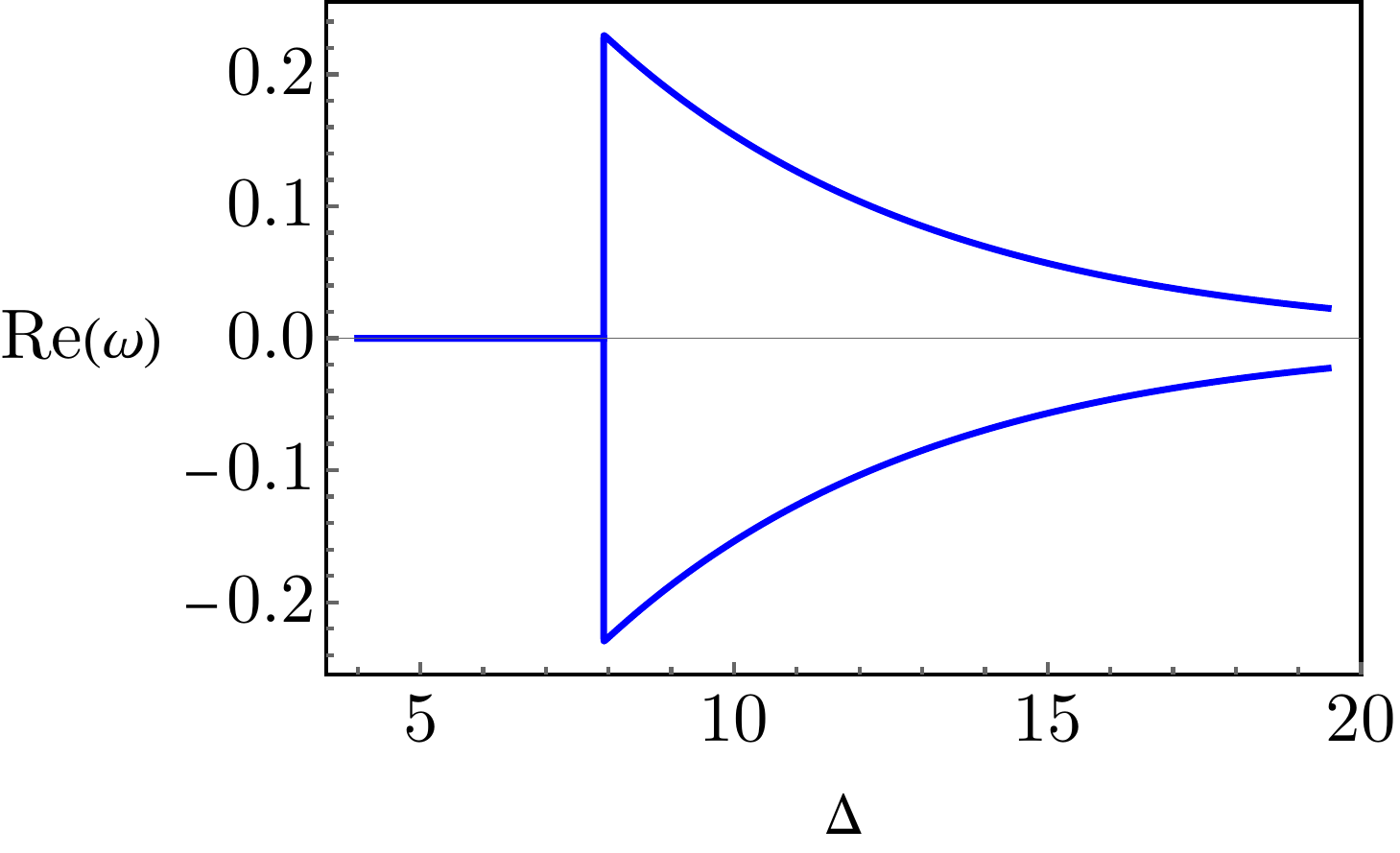}\hfill \includegraphics[width=0.495\textwidth]{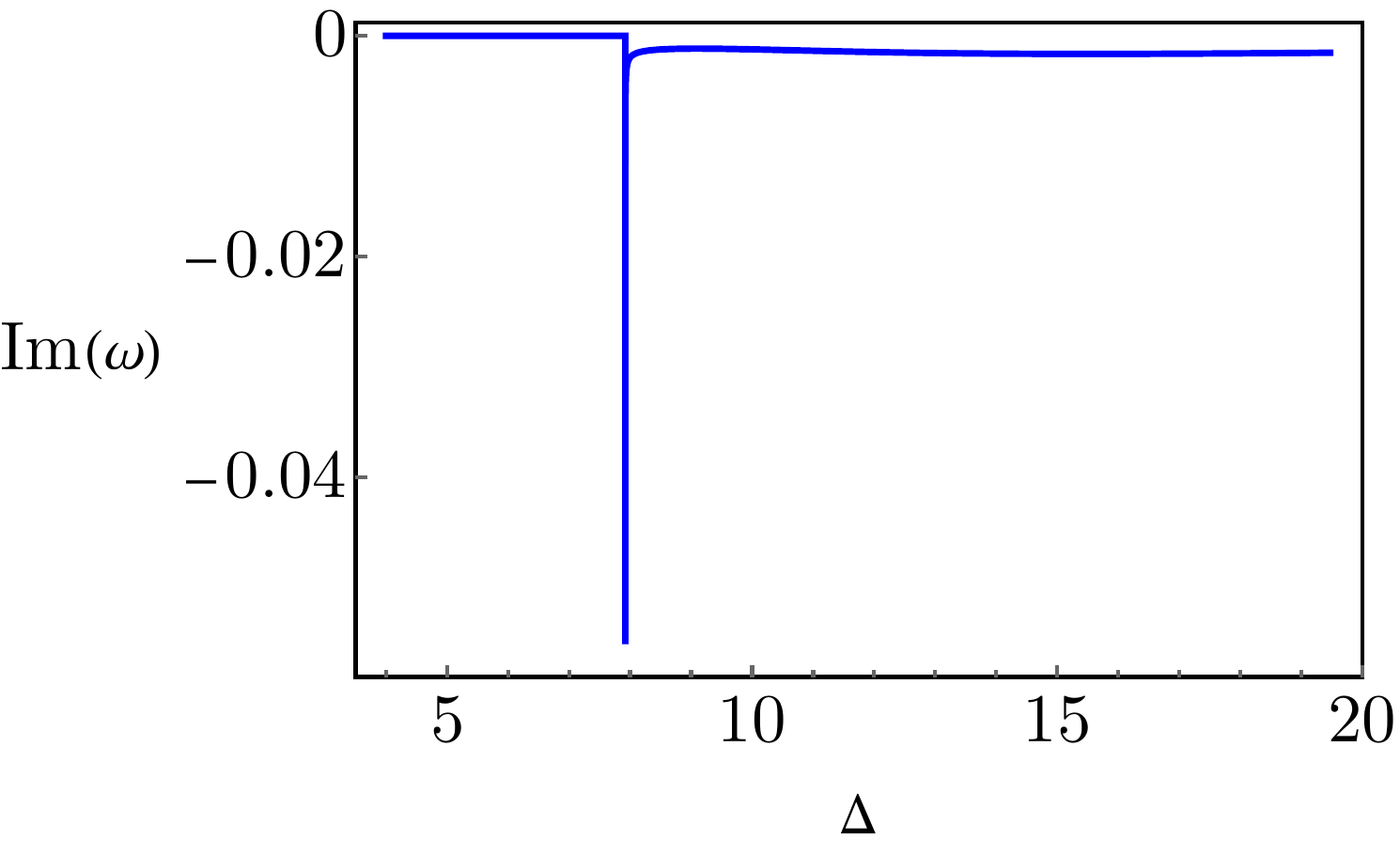}
\caption{\label{gold} Real and imaginary part of the frequency of the lowest lying QNM as a function of the parameter $\Delta$.}
\end{center}
\end{figure}

The structure of the equations of motion here is much more involved than in the Higgs sector since more fields are mutually coupled. However, in the normal phase, it turns out that this sector reduces to a copy of the Higgs sector with extra gauge fields that at low frequencies yield a pole at $\omega=0$. At the phase transition the massive scalar modes hit the origin giving a Goldstone mode and the diffusive modes related to the gauge fields. The Goldstone mode remains gapless, as it should, while the diffusive modes of the mWSM phase yield the complex modes plotted in Fig.~\ref{gold}.
%%%%%%%%%%%%%%%%%%%%%%%%%%%%%%%%%%%%%%%%%%%%%%%%%%%%%%%
%%%%%%%%%%%%%%%%%%%%%%%%%%%%%%%%%%%%%%%%%%%%%%%%%%%%%%%
%%%%%%%%%%%%%%%%%%%%%%%%%%%%%%%%%%%%%%%%%%%%%%%%%%%%%%%%%
%%%%%%%%%%%%%%%%%%%%%%%%%%%%%%%%%%%%%%%%%%%%%%%%%%%

\begin{figure}[t!]
\begin{center}
\includegraphics[width=0.485\textwidth]{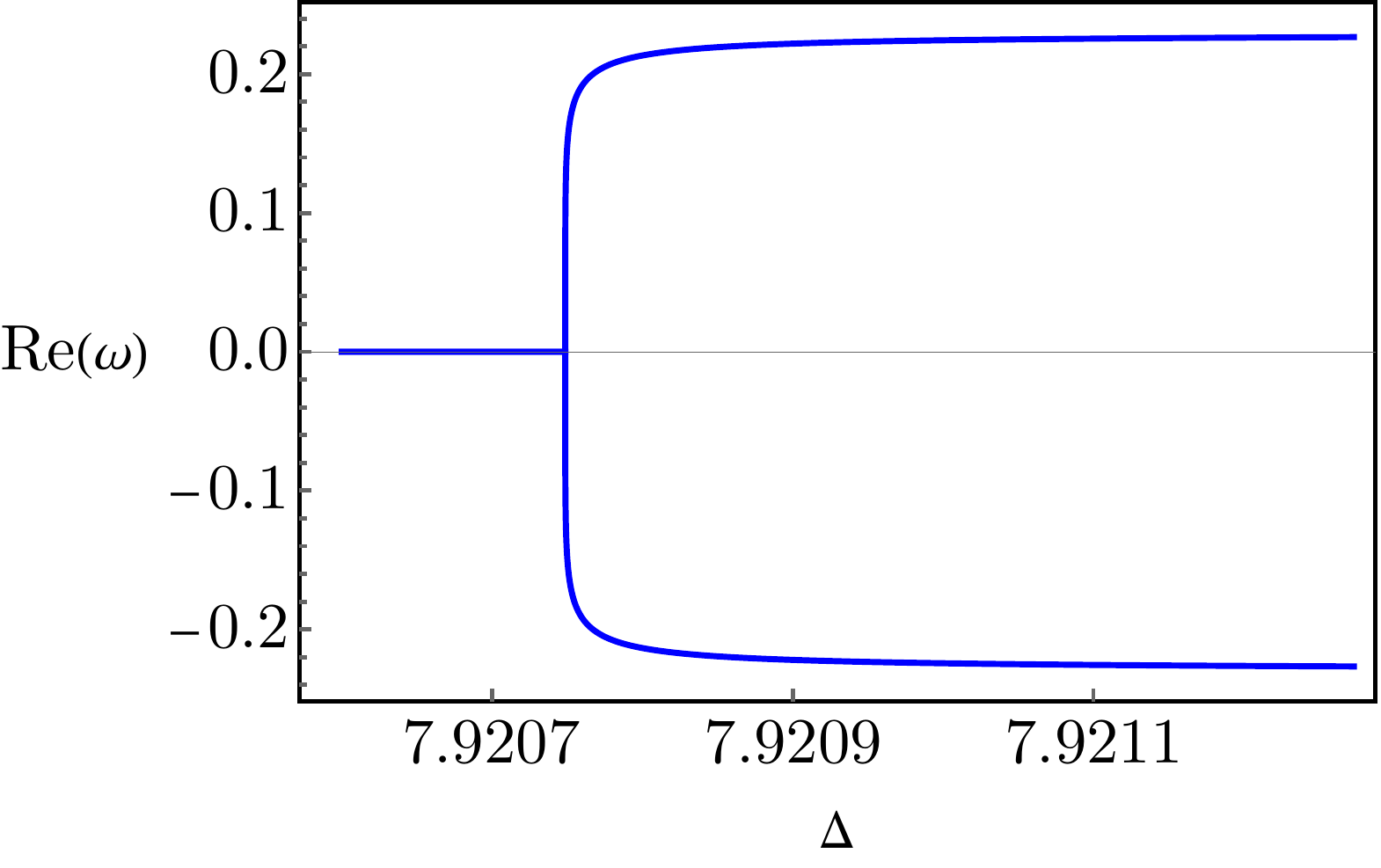}\hfill \includegraphics[width=0.495\textwidth]{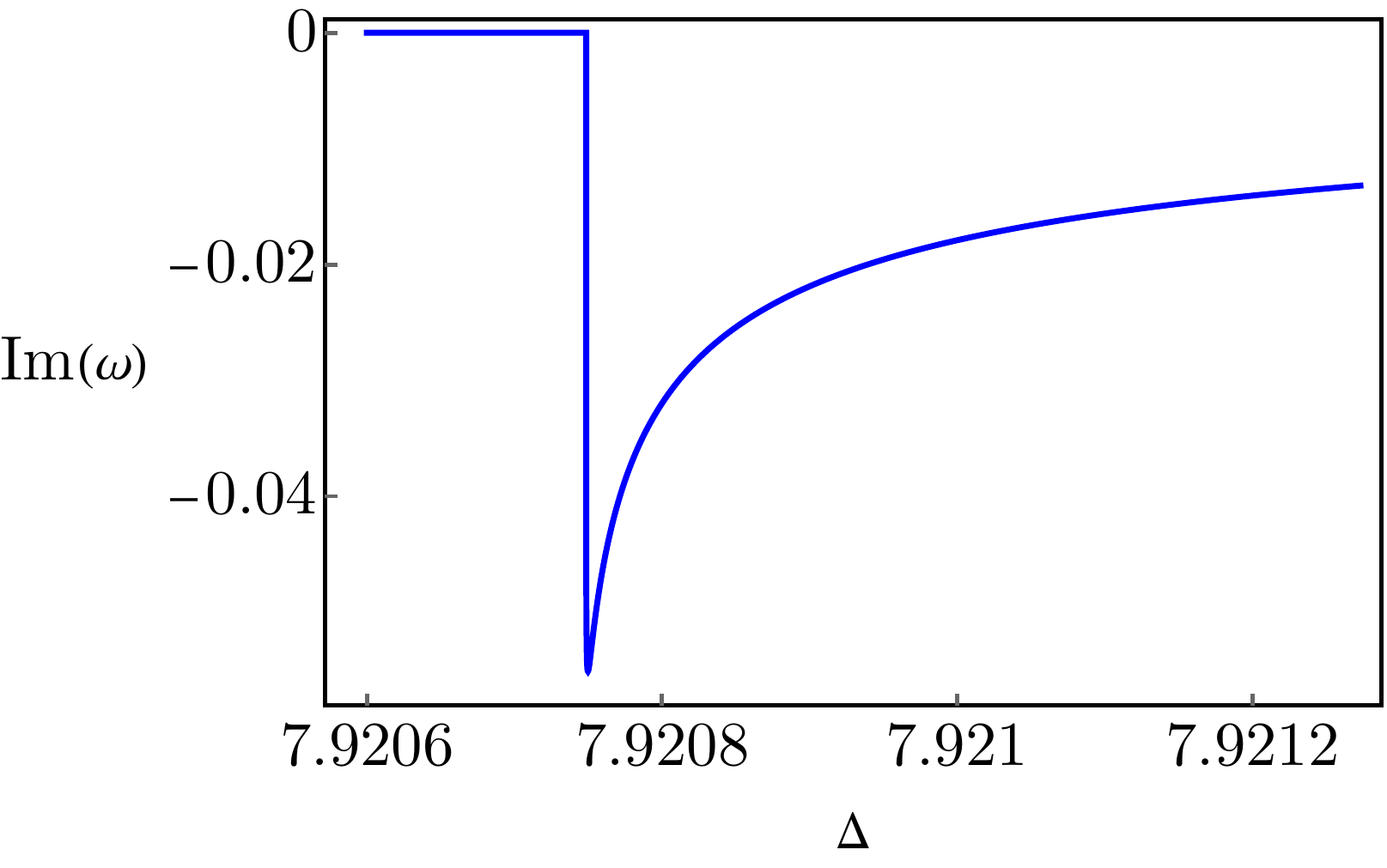}
\caption{\label{goldzi} Real and imaginary part of the lowest lying QNM frequencies a function of $\Delta$.}
\end{center}
\end{figure}

Notice that all the previously discussed features of these modes appear in the vicinity of the phase transition, as shown in Fig.~\ref{gold}, with the region close to the phase transition zoomed in Fig.~\ref{goldzi}. At large $\Delta$ the position of the pole slowly approaches the  origin. The equivalent of these modes for the nematic phase would cross towards the upper half plane, signaling the instability of that phase.

%%%%%%%%%%%%%%%%%%%%%%%%%%%%%%%%%%%%%%%%%%%%%%%%%%%%%
%%%%%%%%%%%%%%%%%%%%%%%%%%%%%%%%%%%%%%%%%%%%%%%%%%%%%
%%%%%%%%%%%%%%%%%%%%%%%%%%%%%%%%%%%%%%%%%%%%%%%%%%
%%%%%%%%%%%%%%%%%%%%%%%%%%%%%%%%%%%%%%%%%%%%%%%%

%%%%%%%%%%%%%%%%%%%%%%%%%%%%%%%%%%%%%%%%%%%%%%555555
%%%%%%%%%%%%%%%%%%%%%%%%%%%%%%%%%%%%%%%%%%%%%%%%%%%%
\section{Anomaly induced transport}
\label{anomtrans}
%%%%%%%%%%%%%%%%%%%%%%%%%%%%%%%%%%%%%%%%%%%%%%%%%%%%%%%%
%%%%%%%%%%%%%%%%%%%%%%%%%%%%%%%%%%%%%%%%%%%%%%%%%%%%%%%%

We now turn to the analysis of the transport coefficients in the {$xy$ nematic} condensate phase. As a first step, we include chemical potentials for the electromagnetic and the $U(1)$ isospins  and generalize the boundary conditions (\ref{eq:boundcond}) to account for these newly introduced sources
\begin{equation}\label{bccm}
 A(r_b) =   \Delta\left(s_x\mathrm dx +s_y\mathrm dy\right)  + x B s_0\mathrm d y+ \left(s_0 \mu+s_3\mu_3\right)dt\,.
\end{equation}
We then also generalize our ansatz for the bulk fields so that it takes the  form
\small
\begin{eqnarray}\label{fullansatz}
 A(r) &=&\left(Q_{x(1)}(r) s_x\,+Q_{x(2)}(r) s_y\,\right) dx +\left( Q_{y(2)}(r)s_y+Q_{y(1)}(r)s_x \right)dy +x B\, s_0\, d y \nonumber\\
 &&+\left(A_{t(0)}(r)s_0+A_{t(3)}(r)s_3\right)dt+\left(A_{z(0)}(r)s_0+A_{z(3)}(r)s_3\right)dz\,.
\end{eqnarray}
\normalsize
These 8 fields are coupled to each other as can be seen from the following equations of motion
\scriptsize
\begin{align}
   & \left(\frac{A_{t(0)}'}{r}\right)'+48 n\lambda B A_{z(0)}'+6n\lambda\left(A_{z(0)}Q_{x(1)}Q_{y(2)}\right)'-6n\lambda \left(A_{z(0)}Q_{y(1)}Q_{x(2)}\right)'=0\nonumber\\
    & \left(\frac{A_{z(0)}'f}{r}\right)'+48 n\lambda B A_{t(0)}'+6n\lambda\left(A_{t(0)}Q_{x(1)}Q_{y(2)}\right)'-6n\lambda \left(A_{t(0)}Q_{y(1)}Q_{x(2)}\right)'=0\nonumber\\
&    \left(\frac{Q_{x(1)}'f}{r}\right)'+\left(\frac{A_{t(3)}^2}{rf}-\frac{A_{z(3)}^2+Q_{y(2)}^2}{r}\right) Q_{x(1)}+24\lambda c(n)\left(A_{z(3)}A_{t(0)}'-A_{z(3)}'A_{t(0)}\right)Q_{y(2)}+\frac{Q_{x(2)}Q_{y(1)}Q_{y(2)}}{r}=0\nonumber\\
 &\left(\frac{Q_{y(2)}'f}{r}\right)'+\left(\frac{A_{t(3)}^2}{rf}-\frac{A_{z(3)}^2+Q_{x(1)}^2}{r}\right) Q_{y(2)}+24\lambda c(n)\left(A_{z(3)}A_{t(0)}'-A_{z(3)}'A_{t(0)}\right)Q_{x(1)}+\frac{Q_{x(2)}Q_{y(1)}Q_{x(1)}}{r}=0\nonumber\\
 &\left(\frac{Q_{x(2)}'f}{r}\right)'+\left(\frac{A_{t(3)}^2}{rf}-\frac{A_{z(3)}^2+Q_{y(1)}^2}{r}\right) Q_{x(2)}+24\lambda c(n)\left(A_{z(3)}A_{t(0)}'-A_{z(3)}'A_{t(0)}\right)Q_{y(1)}+\frac{Q_{y(2)}Q_{y(1)}Q_{x(1)}}{r}=0\nonumber\\
 &\left(\frac{Q_{y(1)}'f}{r}\right)'+\left(\frac{A_{t(3)}^2}{rf}-\frac{A_{z(3)}^2+Q_{x(2)}^2}{r}\right) Q_{y(1)}+24\lambda c(n)\left(A_{z(3)}A_{t(0)}'-A_{z(3)}'A_{t(0)}\right)Q_{x(2)}+\frac{Q_{x(2)}Q_{y(2)}Q_{x(1)}}{r}=0\nonumber\\
 & \left(\frac{A_{z(3)}'f}{r}\right)'+48 n\lambda B A_{t(0)}'+24\lambda c(n)\left(Q_{x(1)}Q_{y(2)}-Q_{x(2)}Q_{y(1)}\right)-\frac{A_{z(3)}|Q|^2}r=0\nonumber\\
  & \left(\frac{A_{t(3)}'}{r}\right)'+48 n\lambda B A_{z(0)}'+24\lambda c(n)\left(Q_{x(1)}Q_{y(2)}-Q_{x(2)}Q_{y(1)}\right)-\frac{A_{t(3)}|Q|^2}r=0.
\end{align}
\normalsize
Here, we have defined $|Q|^2=Q_{x(1)}^2+Q_{x(2)}^2+Q_{y(1)}^2+Q_{y(2)}^2$, and these fields are also subject to the constraint in Eq.~\eqref{constr}.
For the sake of facilitating the numerical computation, we will work exclusively with $n=2$, although we expect a similar qualitative behavior for $n=3$.

\begin{figure}[t!]
\begin{center}
\includegraphics[width=0.495\textwidth]{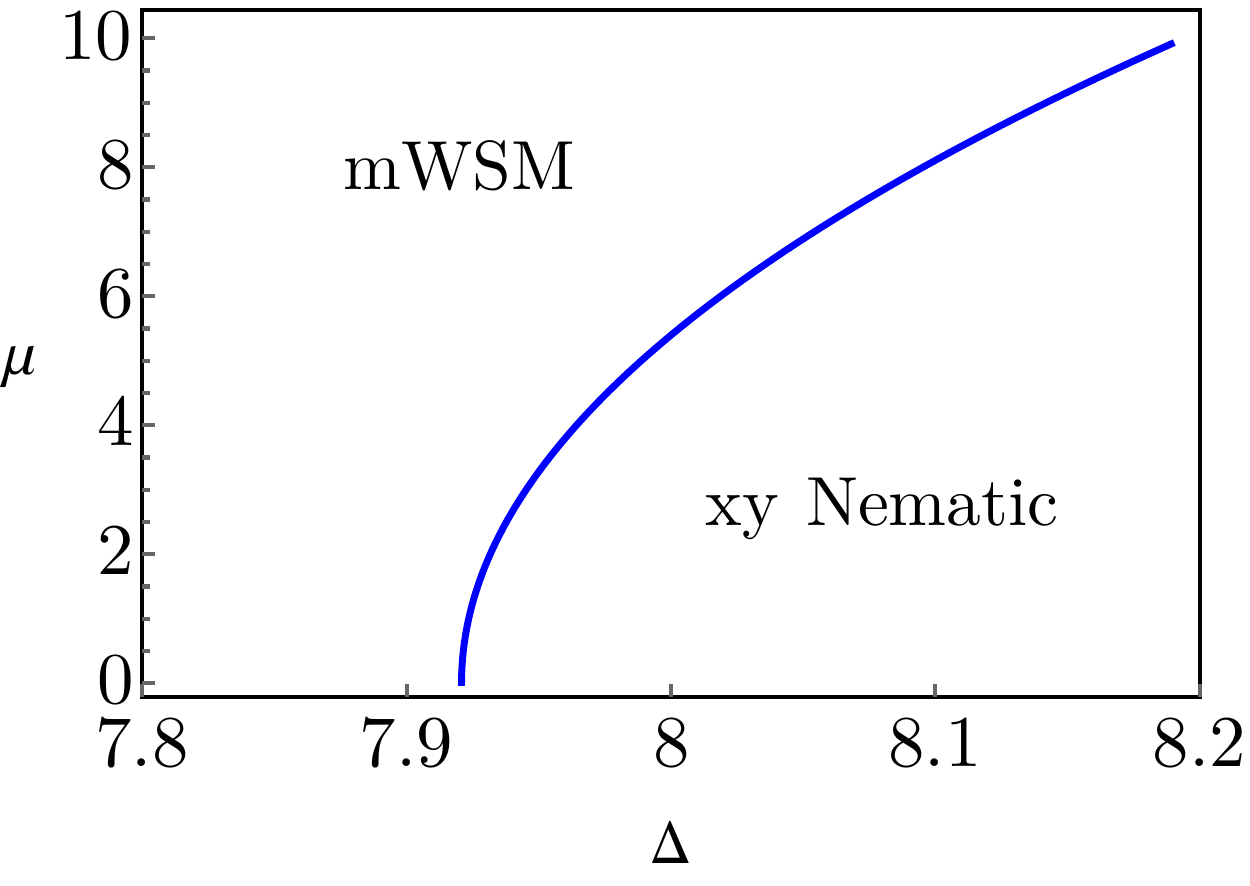}\hfill \includegraphics[width=0.49\textwidth]{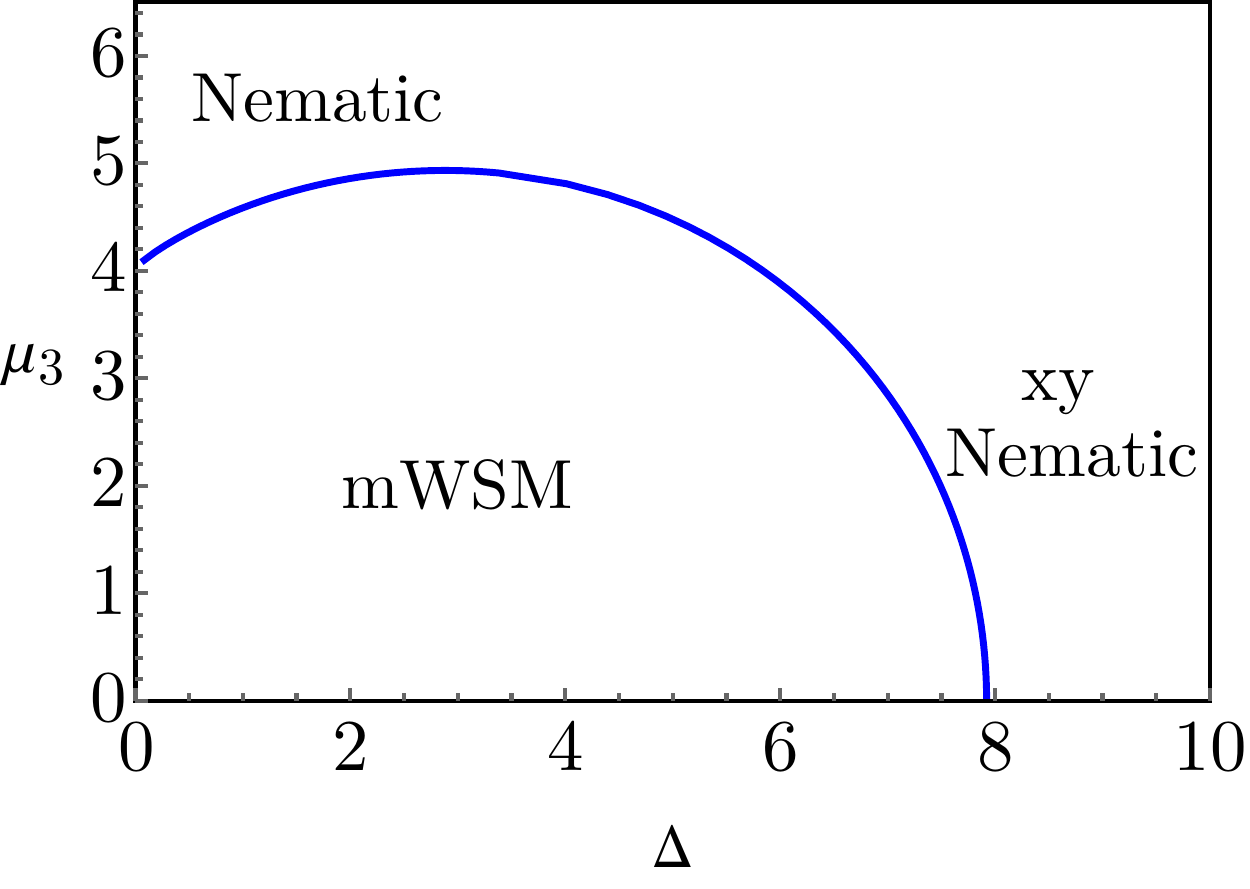}
\caption{\label{phasedi} Phase diagrams at finite $\mu$ (left) and $\mu_3$ (right). The phase boundaries are represented by the blue lines.}
\end{center}
\end{figure}

We analyze  the  phase transition with the abelian and isospin chemical potentials, respectively, $\mu$ and $\mu_3$,  turned on. The results are displayed in Fig.~\ref{phasedi}. In particular, we find that at $\Delta=0$ and $\mu_3$ large enough the nematic phase is favored with respect to the {$xy$ nematic} condensate.
On the other hand, the {$xy$ nematic} condensate is (at $\mu_3=0$) favored at large enough $\Delta$.
We therefore expect that at a certain finite  $\Delta$ there will be a first order phase transition between the two phases. The actual computation of this critical line is quite involved and is beyond the scope of this work.  Furthermore, since we use these sources only to compute the anomalous transport coefficients, the shape of the phase boundary is unimportant as long as there is a finite range of the parameter space where the two phases are stable. 

%%%%%%%%%%%%%%%%%%%%%%%%%%%%%%%%%%%%%%%%%%%%%%%%%%%
%%%%%%%%%%%%%%%%%%%%%%%%%%%%%%%%%%%%%%%%%%%%%%%%%%
%%%%%%%%%%%%%%%%%%%%%%%%%%%%%%%%%%%%%%%%%%%%%%%%%%%%%%%%%%%
%%%%%%%%%%%%%%%%%%%%%%%%%%%%%%%%%%%%%%%%%%%%%%%%%%%%%%%%%%

Now we are ready to study the anomalous transport. We  focus on the effect of the phase transition on the isospin current defined from the UV expansion
\begin{equation}
    A_{z(3)}\approx - J_{z(3)} r^2+\dots
\end{equation}

%%%%%%%%%%%%%%%%%%%%%%%%%%%%%%%%%%%%%%%%%%%%%%%%%

\begin{figure}[t!]
\begin{center}
\includegraphics[width=0.485\textwidth]{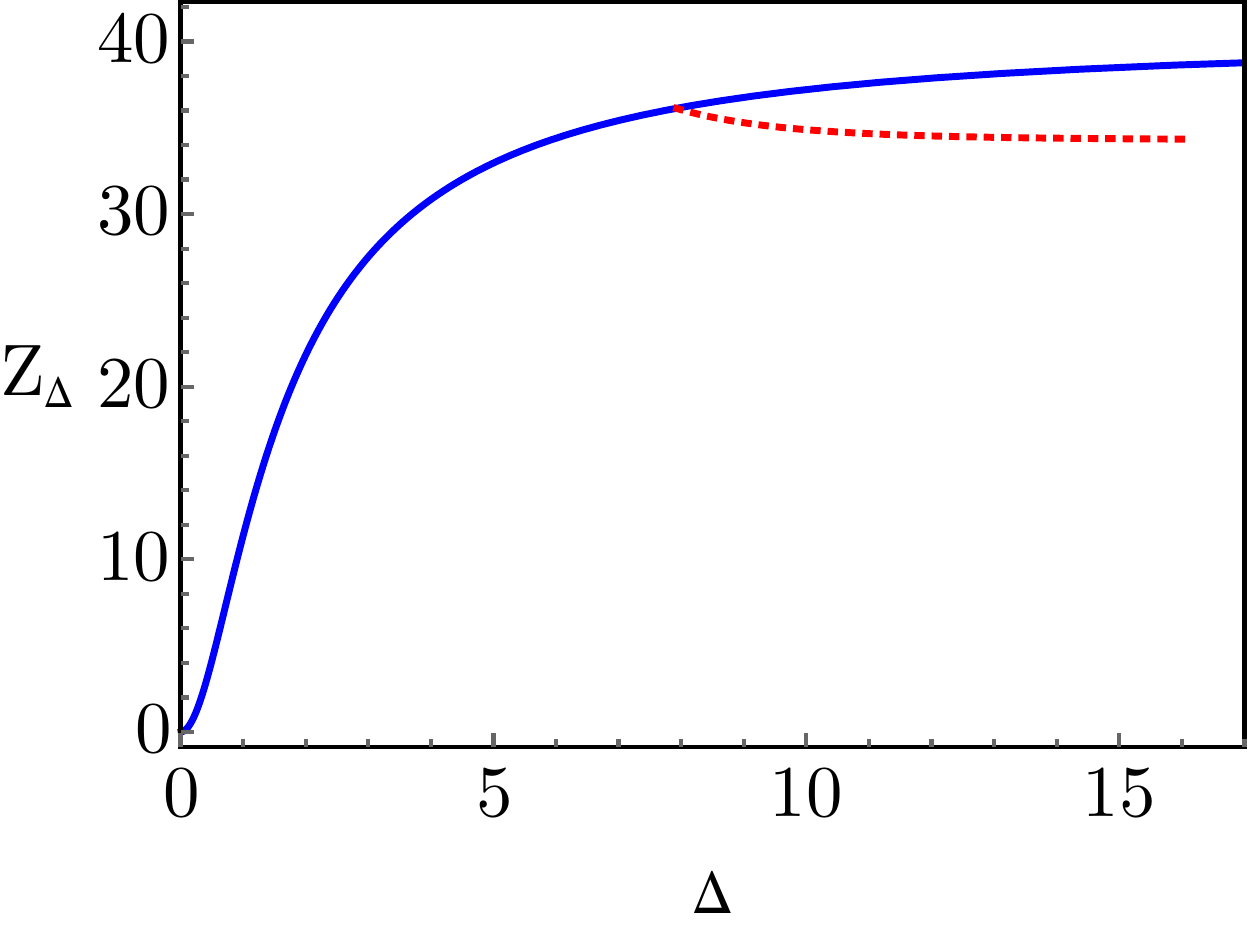}\hfill \includegraphics[width=0.505\textwidth]{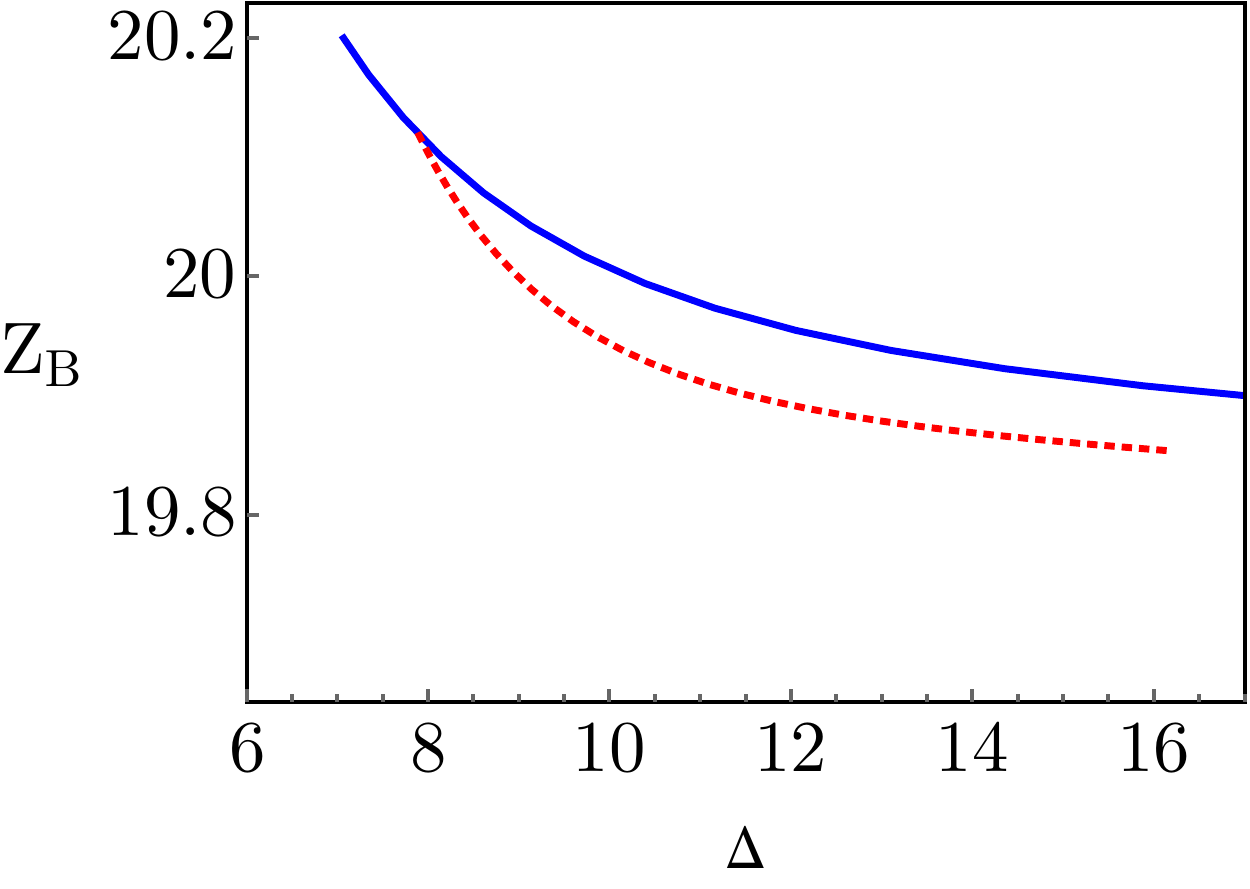}
\caption{\label{ant} Anomalous transport coefficients $Z_\Delta$ (at $\mu=1/2$) and $Z_B$ (at $\mu_3=1/2$) as a function of $\Delta$. The blue (dashed red) line corresponds to the mWSM ({$xy$ nematic} condensate) phase.}
\end{center}
\end{figure}
%%%%%%%%%%%%%%%%%%%%%%%%%%%%%%%%%%%%%%%%%%%%%%%%%%%%%%%%
\noindent In the left panel of Fig.~\ref{ant} we show the isospin current $J_{z(3)}$ computed in the presence of the abelian chemical potential, $\mu$. It is convenient to perform the following  rescaling 
\begin{equation}
    J_{z(3)} =\frac{\mu T^2}{8 \pi^2}Z_{\Delta},
\end{equation}
rendering $Z_{\Delta}$ approximately independent of $\mu$, which holds at least for low enough values of the chemical potential. We see that $Z_{\Delta}\approx \Delta^2$ for small $\Delta$ while it becomes constant for large $\Delta$. Therefore, the effect of the condensate is to deplete the isospin current.

Let us now consider $\mu=0$, $\mu_3\neq0$ and define $Z_B$ as
\begin{equation}
J_{z(3)}=Z_B(\Delta)\frac{c(n)\mu_3}{4 \pi^2}B.
\end{equation}
The isospin chemical potential also leads to the  depletion of isospin carriers, which is consistent with the fact that the {$xy$ nematic} condensate is charged under $s_3$. Finally, as the system is deeper in the condensed phase, the isospin current decreases in comparison with the mWSM phase, see the right panel of Fig.~\ref{ant}.

%%%%%%%%%%%%%%%%%%%%%%%%%%%%%%%%%%%%%%%%%%%%%%%
\section{Discussion and future directions}
%%%%%%%%%%%%%%%%%%%%%%%%%%%%%%%%%%%%%%%%%%%%%%
\label{sec:end}

To summarize, in this paper we studied the phase diagram of the holographic model featuring a mWSM phase~\cite{Dantas:2019rgp}, using a more general ansatz for the bulk fields than for  the mWSM while keeping the same boundary conditions. As the main result, we find that a novel {$xy$ nematic} condensate phase emerges at strong coupling $\Delta$. We then characterize this phase through the lowest lying QNMs and the  anomalous transport coefficients. In the case of the latter, we find a depletion of isospin charge carriers as compared to the mWSM phase.

These results should motivate future work in a few different directions. First of all, 
we would like to point out that the analysis of the QNMs was performed at zero momentum. We expect that analogous computations at finite momenta might be insightful as was the case for the p-wave superfluid~\cite{Herzog:2009ci,Arias:2014msa}. In particular, an analysis of the sound modes in the {$xy$ nematic} phase may reveal some nontrivial features. On the other hand, magnetic field might induce translational symmetry breaking, analogously as  for the p-wave superfluid~\cite{Bu:2012mq}, which is an interesting problem worthwhile pursuing in the case of the {$xy$ nematic} condensate. 

Comparing again with the p-wave superfluids, one might expect to obtain further insights to this system from the entanglement entropy~\cite{Li:2013rhw,Arias:2012py}. To do so, the effects of the back-reaction should be considered, as they encode  properties of the dual stress-energy tensor. 
Related to that, we may study  the gravitational anomaly in this setting, which might leave, on the other hand, a strong imprint on the viscosities, as was recently discussed in the context of holographic WSMs~\cite{Landsteiner:2016stv}. Furthermore, the (shear) viscosity can also probe  the nodal topology~\cite{Moore:2019lay}.

The phase transition into the {$xy$ nematic} condensate should be further characterized in terms of the critical exponents, as was the case for the holographic WSM~\cite{Landsteiner:2015pdh}, and their imprint on various observables. In particular, the optical conductivity displays universal scaling features predicted from the quantum-critical field theory~\cite{ahn2017optical,roy2017}, leading to another open problem to address within the holographic setup. 

We would like to point out that in the model we studied, only one fermion species is included, which  greatly simplifies the analysis,  but is rather artificial, as in general, there are at least two coupled species. To account for this coupling,  adding a second gauge field and to study the associated  anomalies  is an interesting open direction. Furthermore, when considering the interacting left and right fermions, a translational symmetry breaking phase on the lattice corresponds to the opening of the mass gap at the Weyl points at strong coupling, yielding a nontrivial quantum-critical behavior~\cite{roy2017}. It would be important to consider whether a similar phenomenology can be realized in holography. Finally, the lattice realization of the holographic {$xy$ nematic} condensate is also an interesting open problem worthwhile pursuing in the future.

\section*{Acknowledgments}

We would like to thank Nico Grandi {and Francisco Pe\~na-Benitez} for insightful comments. I.S.L. would like to thank IB for hospitality during part of this project. This work was supported in part by the Swedish Research Council, Grant No. VR 2019-04735 (V.J.) and  Fondecyt (Chile) Grant No. 1200399 (R.S.-G.). I.S.L. is a CONICET fellow.

\bibliographystyle{JHEP}
\bibliography{references}{}

\end{document}